\documentclass[english,journal,draftclsnofoot,onecolumn]{IEEEtran}
\usepackage[T1]{fontenc}
\usepackage[latin9]{inputenc}
\usepackage{float}
\usepackage{amsmath}
\usepackage{amssymb}
\usepackage{stackrel}
\usepackage{graphicx}

\makeatletter

\floatstyle{ruled}
\newfloat{algorithm}{tbp}{loa}
\providecommand{\algorithmname}{Algorithm}
\floatname{algorithm}{\protect\algorithmname}

\@ifundefined{date}{}{\date{}}
\@ifundefined{showcaptionsetup}{}{%
 \PassOptionsToPackage{caption=false}{subfig}}
\usepackage{subfig}
\makeatother

\usepackage{babel}
\begin{document}
\title{UNIPOL: Unimodular sequence design via a separable iterative quartic
polynomial optimization for active sensing systems}
\author{Surya Prakash Sankuru, Prabhu Babu, Mohammad Alaee-Kerahroodi}
\maketitle
\begin{abstract}
Sequences having better autocorrelation properties play a crucial
role in enhancing the performance of active sensing systems. Hence,
sequences with good autocorrelation properties are very much in demand.
In this paper, we addressed the problem of designing a unimodular
sequence having better side-lobe levels. We formulated it as a constrained
optimization problem comprising the Integrated Side-lobe Level (ISL)
metric and then proposed an effective algorithm (named UNIPOL - UNImodular
sequence design via a separable iterative POLynomial optimization)
where we perform the polynomial optimization at every iteration. The
novelty of the paper comes from deriving a quartic majorization function
that is separable in the sequence variables and that can be minimized
parallelly. To evaluate the performance of our proposed algorithm
we perform the numerical experiments for different sequence lengths
and confirm that our proposed algorithm is the fastest algorithm to
attain an actual optimum minimizer of the ISL metric. Our proposed
algorithm is also computationally efficient due to its ease of implementation
using the FFT, IFFT operations in a parallel fashion.

\end{abstract}

\section{{Introduction and Problem Formulation}}

One of the major challenges in active sensing systems (RADAR/SONAR)
is to estimate the underlying target parameters which result in the
avoidance of false targets as well as able to discriminate the detected
targets. But, the accuracy of estimated parameters will depend purely
upon the autocorrelation properties of transmit sequences. Hence,
transmit sequences with better autocorrelation properties are always
in demand \cite{2_RadarHandbook}. Apart from the active sensing systems,
some other applications where sequences with good autocorrelation
properties are usually preferred are cryptography and CDMA communication
systems \cite{CDMA}. In practice along with the correlation property,
the aforementioned applications also pose different constraints like
the power (due to its limited quantity available in the system) and
spectral (to avoid the already reserved frequencies) constraints.
Hence, we consider the problem of designing unimodular (constant modulus)
sequences with better autocorrelation properties.

Let $\left\{ x_{n}\right\} _{n=1}^{N}$ be the $N$ length unimodular
sequence having aperiodic auto-correlations $\left\{ r_{k}\right\} _{k=-N+1}^{N-1}$
defined as: 

\begin{equation}
r_{k}=\sum_{n=1}^{N-k}x_{n+k}x_{n}^{*}=r_{-k}^{*},\,k=-N+1,...,N-1.\label{eq:acr}
\end{equation}
Thus the design of a sequence having better side-lobe levels $\left(i.e.\left\{ r_{k}\right\} _{k=1}^{N-1}\right)$
can be formulated as: 

\begin{equation}
\underset{\left|x_{n}\right|=1,\,\forall n}{\text{minimum}}\hphantom{n}\stackrel[k=1]{N-1}{\sum}\left|r_{k}\right|^{2}\label{eq:prob1}
\end{equation}

Some of the existing methods which are capable of generating the unimodular
sequences are given as follows. Earlier, there exist only the analytical
approaches which are able to design the sequences of specific lengths
(for instance, power of $2$). Some of the well known analytical approach
based sequences are the Barker \cite{barker_comm}, \cite{12binary},
Frank \cite{13_Polyphasecode_frank}, Golomb \cite{15_polyphaseseq_Zhang},
Chu \cite{15_polyphaseseq_Zhang}, \cite{3_Radarsignal_Levanon} and
P4 \cite{Impulse_period}, \cite{Set_period} sequences. To overcome
the demarcation of analytical approaches, in the last decade or so,
the computational approaches came into the existence. The authors
in \cite{16_CAN} proposed the CAN algorithm which can solve (\ref{eq:prob1})
by using the alternating minimization technique. However, instead
of solving the problem in (\ref{eq:prob1}) directly, they solved
its approximation. To solve the same ISL minimization problem in (\ref{eq:prob1})
the authors in \cite{17_MISL}, \cite{20_fast_alg_waveform_LI} proposed
the MISL and ISL-NEW algorithms, respectively. Both of them are majorization
minimization (MM) technique based algorithms comprising the linear
(in $x_{n}$) majorization functions which results in slower convergence
issues. The authors in \cite{19_ADMM} used the ADMM technique to
solve an approximation of the problem in (\ref{eq:prob1}). In \cite{CPM_kehrodi}
the authors proposed the CPM algorithm based on the coordinate descent
technique. But, when the length of the sequence increases the CPM
would take a considerable amount of time to converge. Hence, we decided
to propose an efficient technique to solve the problem in (\ref{eq:prob1}).
The major contributions of this letter are: We formulate the problem
of designing a unimodular sequence as a constrained ISL minimization
problem and then proposed a quartic polynomial optimization based
iterative algorithm. In particular, unlike the state-of-the-art MM
based approaches which employ a loose linear upper bound, here we
derive a tighter quartic upper bound for the ISL objective function.
It makes the resultant surrogate minimization problem separable in
the sequence variables which in turn can be solved in a parallel fashion
- this would be an attractive feature if one intends to design sequences
of very large lengths. Later, through computer simulations, we show
the efficacy of our proposed algorithm by comparing it with the state-of-the-art
methods. The rest of the paper is formulated as follows: a brief introduction
of the majorization minimization technique and the proposed algorithm
is given in section II. The numerical simulation results are given
in section III and finally section IV concludes the paper. 

\section{{UNIPOL - The proposed algorithm}}

Before introducing our algorithm, we will discuss the basic principle
of the MM method as follows. The MM method \cite{MM} is a numerical
technique that involves calculating the upper bound (majorization)
function $u(\boldsymbol{\boldsymbol{\boldsymbol{z}}}|\boldsymbol{\boldsymbol{\boldsymbol{z}}}^{k})$
to the original objective function $g(\boldsymbol{z})$ at any point
$\boldsymbol{\boldsymbol{\boldsymbol{z}}}^{k}$ and then minimizing
the upper bound to arrive at the next iterate point. The constructed
majorization function has to satisfy the following properties.

\begin{equation}
u(\boldsymbol{\boldsymbol{\boldsymbol{z}}}^{k}|\boldsymbol{\boldsymbol{\boldsymbol{z}}}^{k})=g(\boldsymbol{\boldsymbol{\boldsymbol{z}}}^{k})\text{ and }u(\boldsymbol{\boldsymbol{z}}|\boldsymbol{\boldsymbol{\boldsymbol{z}}}^{k})\geq g(\boldsymbol{\boldsymbol{z}}),\:\forall\boldsymbol{\boldsymbol{z}}\in\boldsymbol{\boldsymbol{\boldsymbol{Z}}}.\label{eq:6-1}
\end{equation}
where $\boldsymbol{\boldsymbol{\boldsymbol{Z}}}$ is a feasible set.
As the MM technique is an iterative process, it will generate the
sequence of points $\left\{ \boldsymbol{\boldsymbol{z}}\right\} =\boldsymbol{\boldsymbol{\boldsymbol{z}}}^{1},\boldsymbol{\boldsymbol{\boldsymbol{z}}}^{2},\boldsymbol{\boldsymbol{\boldsymbol{z}}}^{3},.....,\boldsymbol{\boldsymbol{\boldsymbol{z}}}^{m}$
according to the following update rule: 

\begin{equation}
\boldsymbol{\boldsymbol{\boldsymbol{z}}}^{k+1}\triangleq\text{arg}\min_{\boldsymbol{\boldsymbol{z}\in\boldsymbol{\boldsymbol{\boldsymbol{Z}}}}}u(\boldsymbol{\boldsymbol{z}}|\boldsymbol{\boldsymbol{\boldsymbol{z}}}^{k}).\label{eq:7-1}
\end{equation}

The value of the objective function computed at every iterate point
(\ref{eq:7-1}) will satisfy the monotonic decreasing property, i.e.
\begin{equation}
g(\boldsymbol{\boldsymbol{\boldsymbol{z}}}^{k+1})\leq u(\boldsymbol{\boldsymbol{\boldsymbol{z}}}^{k+1}|\boldsymbol{\boldsymbol{\boldsymbol{z}}}^{k})\leq u(\boldsymbol{\boldsymbol{\boldsymbol{z}}}^{k}|\boldsymbol{\boldsymbol{\boldsymbol{z}}}^{k})=g(\boldsymbol{\boldsymbol{\boldsymbol{z}}}^{k}).\label{eq:MM-1}
\end{equation}

Now, we will start with our problem of interest in (\ref{eq:prob1}).
Here, the cost function is expressed in terms of the autocorrelation
values, but by re-expressing it in frequency domain (using the Parseval's
theorem), we arrive at the following equivalent form (a short proof
of the same can be found in the appendix of \cite{16_CAN}): 

\begin{equation}
\stackrel[k=1]{N-1}{\sum}\left|r_{k}\right|^{2}\triangleq\frac{1}{2N}\stackrel[p=1]{2N}{\sum}\left|\left|\stackrel[n=1]{N}{\sum}x_{n}e^{-j\omega_{p}n}\right|^{2}-N\right|^{2}\label{eq:parseval}
\end{equation}

where $\omega_{p}=\frac{2\pi}{2N}p$.

By expanding the square in the equivalent objective function and using
the fact that the energy of the signal is constant (i.e. $\stackrel[p=1]{2N}{\sum}\left|\stackrel[n=1]{N}{\sum}x_{n}e^{-j\omega_{p}n}\right|^{2}=2N^{2}$),
the ISL minimization problem can be expressed as: 

\begin{equation}
\underset{\left|x_{n}\right|=1,\,\forall n}{\text{minimum}}\hphantom{n}\stackrel[p=1]{2N}{\sum}\left|\stackrel[n=1]{N}{\sum}x_{n}e^{-j\omega_{p}n}\right|^{4}\label{eq:prob2}
\end{equation}

Now by using the Jensen's inequality (see example-$11$ in \cite{MM}),
at any given $x_{n}^{t}$, the cost function in the above problem
can be majorized (or) upper bounded as follows: 

\begin{equation}
\begin{array}{ll}
\stackrel[p=1]{2N}{\sum}\left|\stackrel[n=1]{N}{\sum}x_{n}e^{-j\omega_{p}n}\right|^{4}\\
\quad\leq\stackrel[p=1]{2N}{\sum}\stackrel[n=1]{N}{\sum}\frac{1}{N}\left|N\left(x_{n}e^{-j\omega_{p}n}-x_{n}^{t}e^{-j\omega_{p}n}\right)+\stackrel[n^{'}=1]{N}{\sum}x_{n^{'}}^{t}e^{-j\omega_{p}n^{'}}\right|^{4}\label{eq:maj1}
\end{array}
\end{equation}

One can observe that both the upper bound and exact functions are
quartic in nature. Hence, it confirms that by using Jensen's inequality
we are able to construct a quartic majorization function to the original
quartic objective function and itself is the novelty of our paper.
After pulling out the factor $N$ and the complex exponential in the
first and second terms we arrive at the following equivalent upper
bound function: 
\begin{equation}
u(x_{n}|x_{n}^{t})=\stackrel[p=1]{2N}{\sum}\stackrel[n=1]{N}{\sum}\left|x_{n}-x_{n}^{t}+\frac{1}{N}\stackrel[n^{'}=1]{N}{\sum}x_{n^{'}}^{t}e^{-j\omega_{p}\left(n^{'}-n\right)}\right|^{4}\label{eq:m2}
\end{equation}
It's worth pointing out that the above surrogate function is separable
in $x_{n}$ and a generic function (independent of $n$) will be given
by: 

\begin{equation}
u({x}|x^{t})=\stackrel[p=1]{2N}{\sum}\left|x-x^{t}+\frac{1}{N}\stackrel[n^{'}=1]{N}{\sum}x_{n^{'}}^{t}e^{-j\omega_{p}\left(n^{'}-q\right)}\right|^{4}\label{eq:m3}
\end{equation}

where $q$ denotes the corresponding variable index of $x$. The surrogate
in (\ref{eq:m3}) can be rewritten more compactly as: 

\begin{equation}
u({x}|x^{t})=\stackrel[p=1]{2N}{\sum}\left|x-\alpha_{p}\right|^{4}\label{eq:m4}
\end{equation}

where $\alpha_{p}=x^{t}-\frac{1}{N}\stackrel[n^{'}=1]{N}{\sum}x_{n^{'}}^{t}e^{-j\omega_{p}\left(n^{'}-q\right)}.$

So, any individual surrogate minimization problem would be as follows: 

\begin{equation}
\underset{\left|x\right|=1}{\text{min}}\hphantom{n}\stackrel[p=1]{2N}{\sum}\left|x-\alpha_{p}\right|^{4}\label{eq:prob3}
\end{equation}

where $\alpha_{p}$ is a complex variable with $\left|\alpha_{p}\right|\neq1$.
The cost function in the above problem (\ref{eq:prob3}) can be rewritten
as: 

\begin{equation}
\begin{aligned}\stackrel[p=1]{2N}{\sum}\left|x-\alpha_{p}\right|^{4} & =\stackrel[p=1]{2N}{\sum}\left(\left|x-\alpha_{p}\right|^{2}\right)^{2}=\stackrel[p=1]{2N}{\sum}\left(1-2\text{Re}\left(\alpha_{p}^{*}x\right)+\left|\alpha_{p}\right|^{2}\right)^{2}\\
 & =\stackrel[p=1]{2N}{\sum}\left(4\left(\text{Re}\left(\alpha_{p}^{*}x\right)\right)^{2}-4\text{Re}\left(\alpha_{p}^{*}x\right)\left(1+\left|\alpha_{p}\right|^{2}\right)+\text{const}\right)
\end{aligned}
\label{eq:m5}
\end{equation}

By neglecting the constant terms, (\ref{eq:m5}) simplifies as: 

\begin{equation}
=\stackrel[p=1]{2N}{\sum}\left(\left(\alpha_{p}^{*}x+x^{*}\alpha_{p}\right)^{2}-4\text{Re}\left(\alpha_{p}^{*}x\right)\left(1+\left|\alpha_{p}\right|^{2}\right)\right)\label{eq:m6}
\end{equation}

\begin{equation}
=\stackrel[p=1]{2N}{\sum}\left(\left[\left(\alpha_{p}^{*}\right)^{2}x^{2}+\left(x^{*}\right)^{2}\alpha_{p}^{2}\right]-4\text{Re}\left(\alpha_{p}^{*}x\right)\left(1+\left|\alpha_{p}\right|^{2}\right)+\text{const}\right)\label{eq:m7}
\end{equation}

By neglecting the constant terms, (\ref{eq:m7}) can be rewritten
as: 

\begin{equation}
\begin{aligned} & =\stackrel[p=1]{2N}{\sum}\left(2\text{Re}\left(\left(\alpha_{p}^{*}\right)^{2}x^{2}\right)-4\text{Re}\left(\alpha_{p}^{*}x\right)\left(1+\left|\alpha_{p}\right|^{2}\right)\right)\\
 & =\text{Re}\left(ax^{2}\right)-\text{Re}\left(bx\right)
\end{aligned}
\label{eq:m8}
\end{equation}

where $a=\stackrel[p=1]{2N}{\sum}2\left(\alpha_{p}^{*}\right)^{2}$
and $b=\stackrel[p=1]{2N}{\sum}4\alpha_{p}^{*}\left(1+\left|\alpha_{p}\right|^{2}\right)$. 

So, the final surrogate minimization problem came to be: 

\begin{equation}
\underset{\left|x\right|=1}{\text{min}}\hphantom{n}\text{Re}\left(ax^{2}-bx\right)\label{eq:prob4}
\end{equation}

Although the above problem looks like a simple univariable optimization
problem, it doesn't have any closed form solution. So, to compute
its minimizer we express the constraint $\left|x\right|=1$ as $x=e^{j\theta}$
and then the first order KKT condition of the above problem is given
as: 
\begin{equation}
\frac{d}{d\theta}(\text{Re}\left(ae^{2j\theta}-be^{j\theta}\right))=\text{Re}\left(2aje^{2j\theta}-jbe^{j\theta}\right)=0.\label{eq:real}
\end{equation}
By defining $2ja=-2a_{I}+j2a_{R}$ and $jb=-b_{I}+jb_{R}$, where
$a_{R},b_{R},a_{I},b_{I}$ are the real and imaginary parts of $a$
and $b$, respectively, then the KKT condition becomes as:
\begin{equation}
\begin{array}{ll}
\text{Re}\left(2aje^{2j\theta}-jbe^{j\theta}\right)\\
\:=-2a_{I}\cos\left(2\theta\right)-2a_{R}\sin\left(2\theta\right)+b_{I}\cos\left(\theta\right)+b_{R}\sin\left(\theta\right)=0
\end{array}
\end{equation}
which further simplifies as: 
\begin{equation}
2a_{I}\cos\left(2\theta\right)+2a_{R}\sin\left(2\theta\right)-b_{I}\cos\left(\theta\right)-b_{R}\sin\left(\theta\right)=0
\end{equation}
Now, if we let $\beta=tan\left(\frac{\theta}{2}\right)$, then $sin\left(\theta\right)=\frac{2\beta}{1+\beta^{2}}$,
$cos\left(\theta\right)=\frac{1-\beta^{2}}{1+\beta^{2}}$, $sin\left(2\theta\right)=\frac{4\beta\left(1-\beta^{2}\right)}{\left(1+\beta^{2}\right)^{2}}$,
$cos\left(2\theta\right)=\frac{1+\beta^{4}-6\beta^{2}}{\left(1+\beta^{2}\right)^{2}}$,
then the KKT condition can be rewritten as: 

\begin{equation}
\frac{2a_{I}\left(1+\beta^{4}-6\beta^{2}\right)}{\left(1+\beta^{2}\right)^{2}}+\frac{2a_{R}\left(4\beta\left(1-\beta^{2}\right)\right)}{\left(1+\beta^{2}\right)^{2}}-\frac{b_{I}\left(1-\beta^{2}\right)}{1+\beta^{2}}-\frac{b_{R}2\beta}{1+\beta^{2}}=0\label{eq:prob6}
\end{equation}
\begin{equation}
\frac{2a_{I}(1+\beta^{4})-12a_{I}\beta^{2}+8a_{R}(\beta-\beta^{3})-b_{I}(1-\beta^{4})-2b_{R}(\beta+\beta^{3})}{\left(1+\beta^{2}\right)^{2}}=0\label{eq:prob7}
\end{equation}
which can be rewritten as:
\begin{equation}
\frac{p_{4}\beta^{4}+p_{3}\beta^{3}+p_{2}\beta^{2}+p_{1}\beta+p_{0}}{\left(1+\beta^{2}\right)^{2}}=0\label{eq:prob8}
\end{equation}
with
\begin{equation}
\begin{aligned}p_{4} & =2a_{I}+b_{I},\:p_{3}=-8a_{R}-2b_{R}\\
p_{2} & =-12a_{I},\:p_{1}=8a_{R}-2b_{R},\:p_{0}=2a_{I}-b_{I}
\end{aligned}
\label{eq:constants}
\end{equation}
Since $(1+\beta^{2})\neq0$, (\ref{eq:prob8}) is equivalent to: 
\begin{equation}
p_{4}\beta^{4}+p_{3}\beta^{3}+p_{2}\beta^{2}+p_{1}\beta+p_{0}=0
\end{equation}
which is a quartic polynomial. So, we calculate the roots of this
quartic polynomial and choose the root which gives the least value
of objective in (\ref{eq:prob4}) as the minimizer of surrogate minimization
problem. The pseudocode of the proposed algorithm is given in Algorithm-1. 

Some remarks on the proposed algorithm are given below: 
\begin{itemize}
\item The major chunk of the computational complexity comes from the calculation
of $\alpha_{p}$'s, which can be done easily via the FFT operations
($i.e.,$ the second term in $\alpha_{p}$ can be written as $\frac{1}{N}\stackrel[n^{'}=1]{N}{\sum}x_{n^{'}}^{t}e^{-j\omega_{p}n^{'}}e^{j\omega_{p}q}$
where $\stackrel[n^{'}=1]{N}{\sum}x_{n^{'}}^{t}e^{-j\omega_{p}n^{'}}$
is FFT of $\boldsymbol{\boldsymbol{x}}$). Hence, the computational
complexity of UNIPOL came to be $\mathcal{O}\left(N\,logN\right)$
and the space complexity as $\mathcal{O}\left(N\right)$. The computational
complexity of state-of-the-art methods (which we compare in the numerical
section) is given as: CAN-$\mathcal{O}(N\,log\,N)$, MISL-$\mathcal{O}(N\,log\,N)$,
ISL-NEW-$\mathcal{O}(N\,log\,N)$, ADMM-$\mathcal{O}(N^{3})$, CPM-$\mathcal{O}(K\,log\,N)$,
where $K\text{ }$denotes the number of iterations incurred to implement
the bisection method. We noticed that our proposed algorithm is having
either same or better computational complexity than the state-of-the-art
algorithms. 
\item It's worth pointing that the original ISL objective function is quartic
in $x_{n}$ and the upper bounds employed by the methods like MISL
and ISL-NEW are linear in $x_{n}$, on the other hand, the surrogate
function derived in our algorithm (see eq (\ref{eq:prob3})) is quartic
in $x_{n}$ . Thus, our surrogate function would be tighter upper
bound for ISL function than the surrogates of MISL and ISL-NEW algorithms.
As we know that the convergence of MM algorithms will depend mostly
on the tightness of surrogate function, we expect the convergence
of our algorithm would be much better than the MISL and ISL-NEW algorithms.
Indeed, we will show in the numerical simulations section that our
proposed algorithm is faster than the MISL and ISL-NEW algorithms. 
\end{itemize}
\begin{algorithm}[H]
\textbf{Require}: sequence length $\text{\textquoteleft}N\text{\textquoteright}$

1:set $t=0$, initialize $\boldsymbol{x}^{0}$

2:\textbf{ repeat}

3:$\hphantom{nn}$set $n=1$

4:$\hphantom{nnnn}$\textbf{ repeat}

5:$\hphantom{nnnnn}$compute $\alpha_{p},\,\forall p$.

6:$\hphantom{nnnnn}$compute $a,b.$

7:$\hphantom{nnnnn}$compute $p_{0},p_{1},p_{2},p_{3},p_{4}$ using
(\ref{eq:constants}).

8:$\hphantom{nnnnn}$compute optimal $\beta_{n}$ by solving the KKT
condition.

9:$\hphantom{nnnnn}$$n\longleftarrow n+1$

10:$\hphantom{nnn}$\textbf{until $n=N$}

11:$\hphantom{n}$$x_{n}^{t+1}=e^{j2\:\textrm{arc tan}\left(\beta_{n}\right)},\,\forall n.$

12:$\hphantom{n}$ $t\longleftarrow t+1$

13:\textbf{ until }convergence \caption{:UNIPOL}
\end{algorithm}

\begin{itemize}
\item In the development of our algorithm, we derive a surrogate function
which is separable in the sequence variables and shows the minimization
of it only over a generic variable which is identical for all the
variables. So, our algorithm can be implemented parallelly when we
solve the $N$ surrogate minimization problems (the steps $3-10$
in the UNIPOL algorithm can be implemented parallelly). This is an
attractive feature that would be handy when we design sequences of
larger lengths $N\approx10^{6}$. 
\item Since our algorithm is derived based on the MM principle, it enjoys
the monotonic descent property and the standard convergence proofs
would be applicable to establish the convergence of our algorithm,
please see section IIc in \cite{MM} for more details on convergence
of MM steps. 
\end{itemize}

\section{{Numerical Simulations}}

In this section, we numerically evaluate the performance of our proposed
algorithm. The numerical simulations are conducted for different sequence
lengths like $N=50,100,225,400,625,900,1000,1300$. For each length,
the evolution of ISL value with respect to the iterations and convergence
time, autocorrelation properties of the converged sequences are computed.
The algorithms chosen for comparison are CAN, MISL, ISL-NEW, ADMM, 

\begin{figure}[H]
\subfloat[ISL vs iterations for $N=100$]{\includegraphics[scale=0.55]{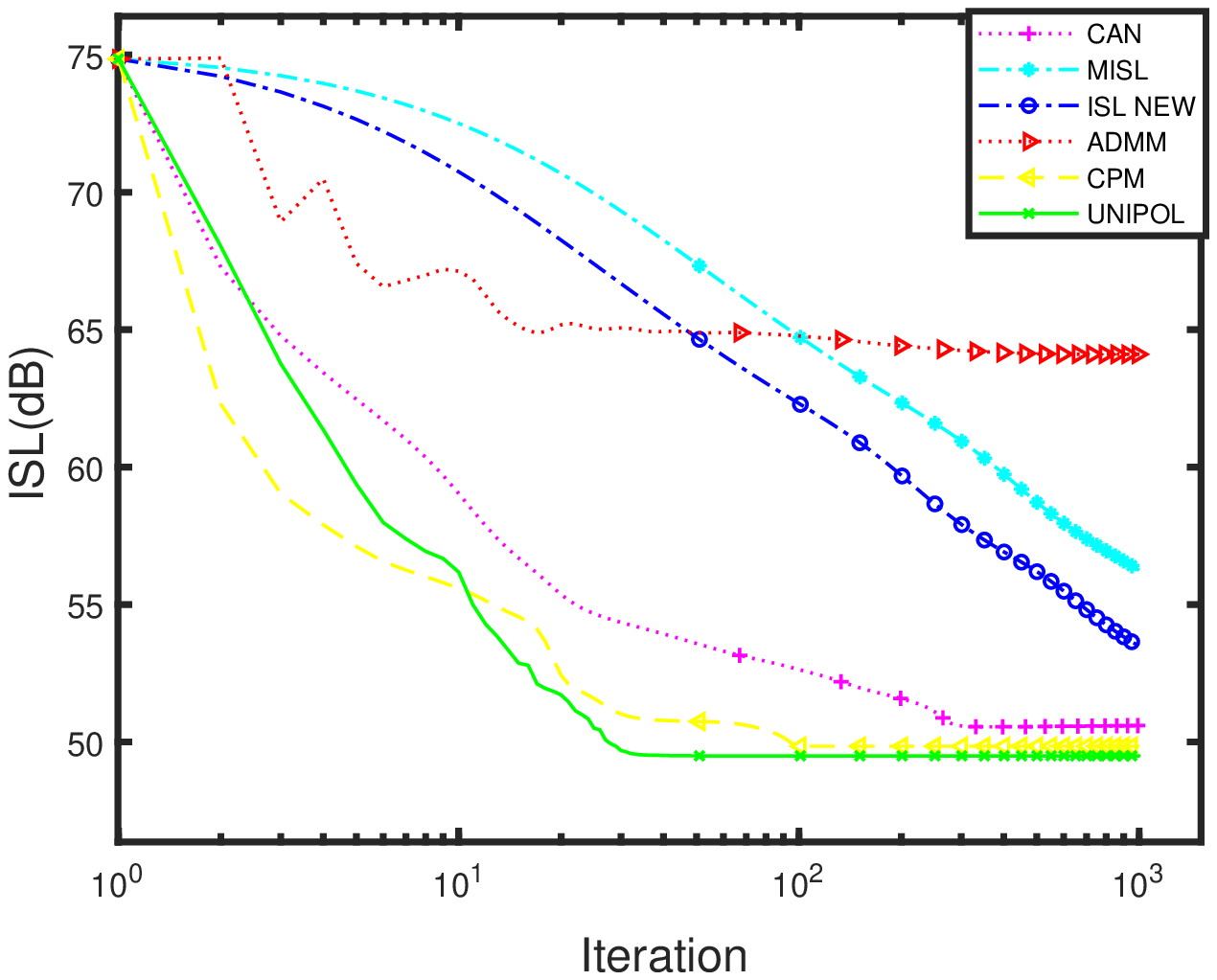}

}\subfloat[ISL vs iterations for $N=1000$]{\includegraphics[scale=0.55]{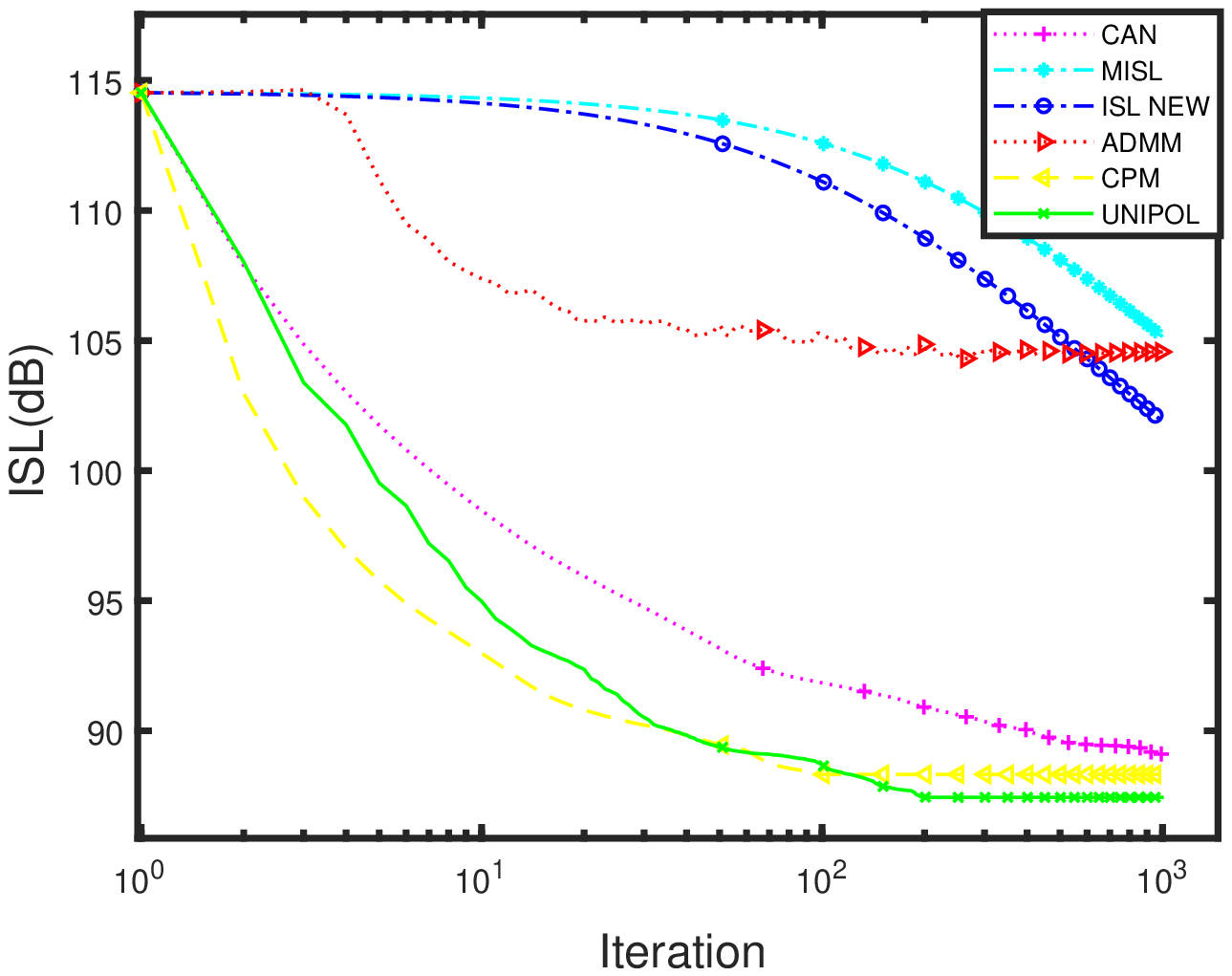}

}

\subfloat[ISL vs time for $N=100$]{\includegraphics[scale=0.55]{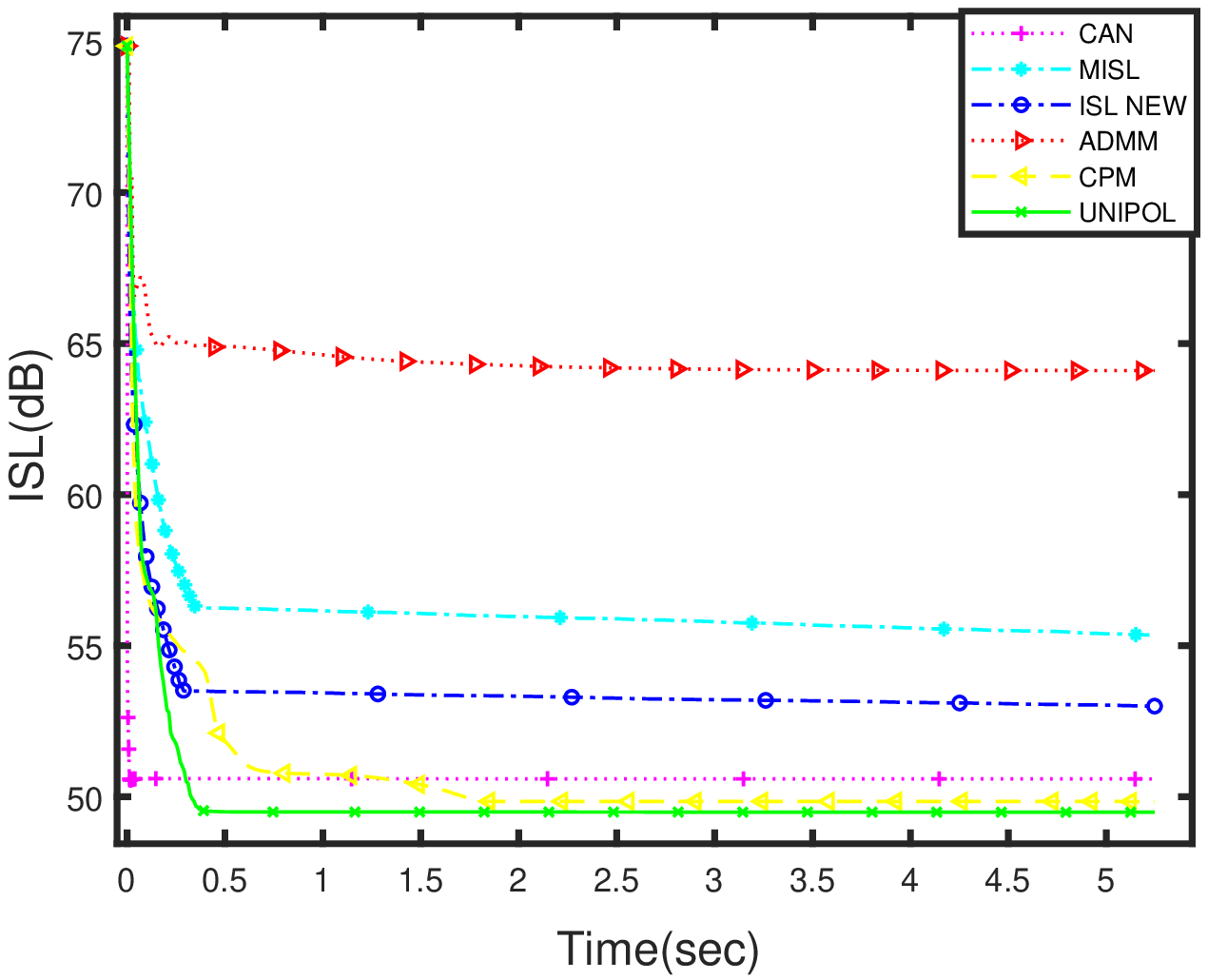}

}\subfloat[ISL vs time for $N=1000$]{\includegraphics[scale=0.55]{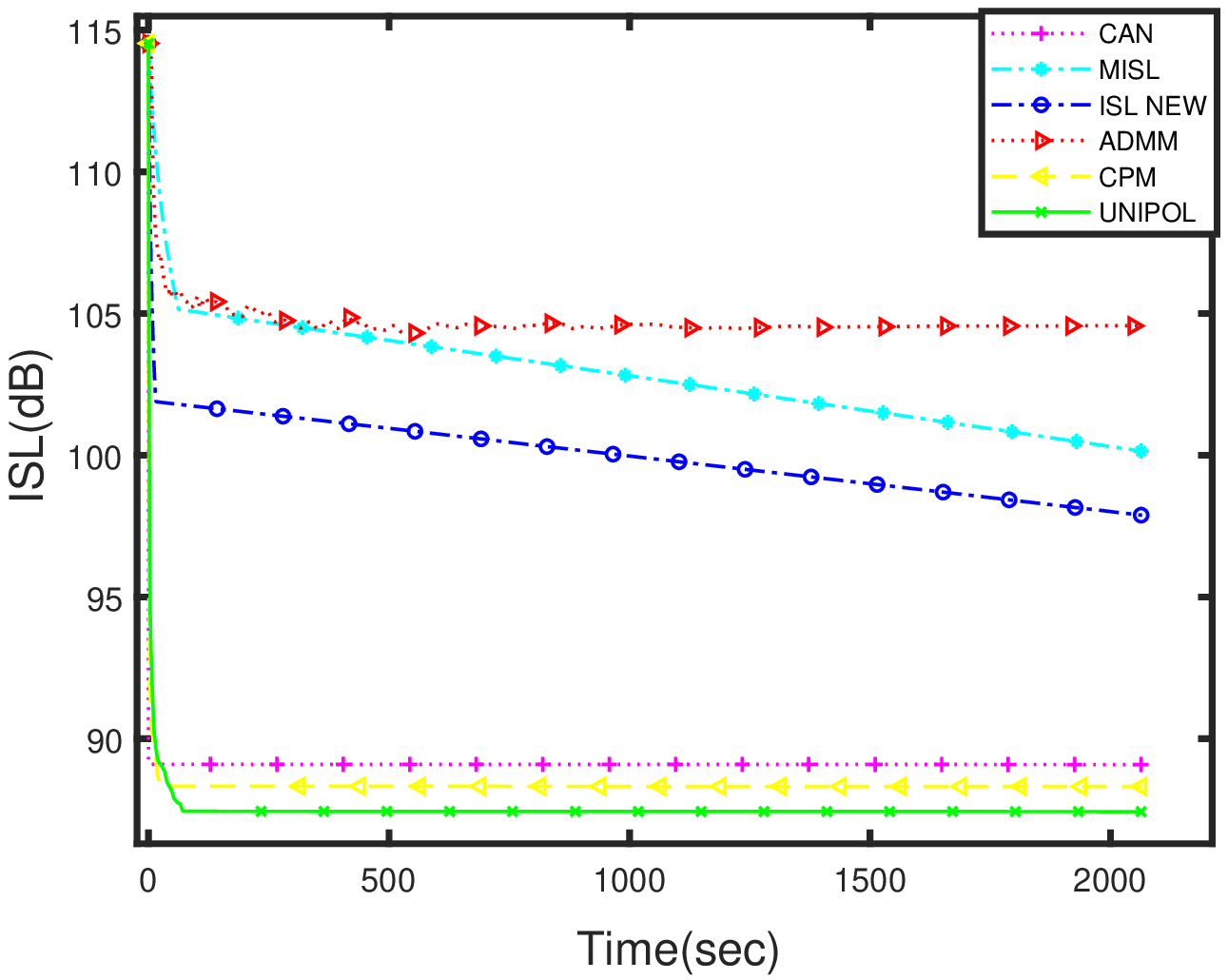}

}

\subfloat[$20\log10\left(\left|r_{k}\right|/N\right)$ vs $k$ for $N=100$]{\includegraphics[scale=0.55]{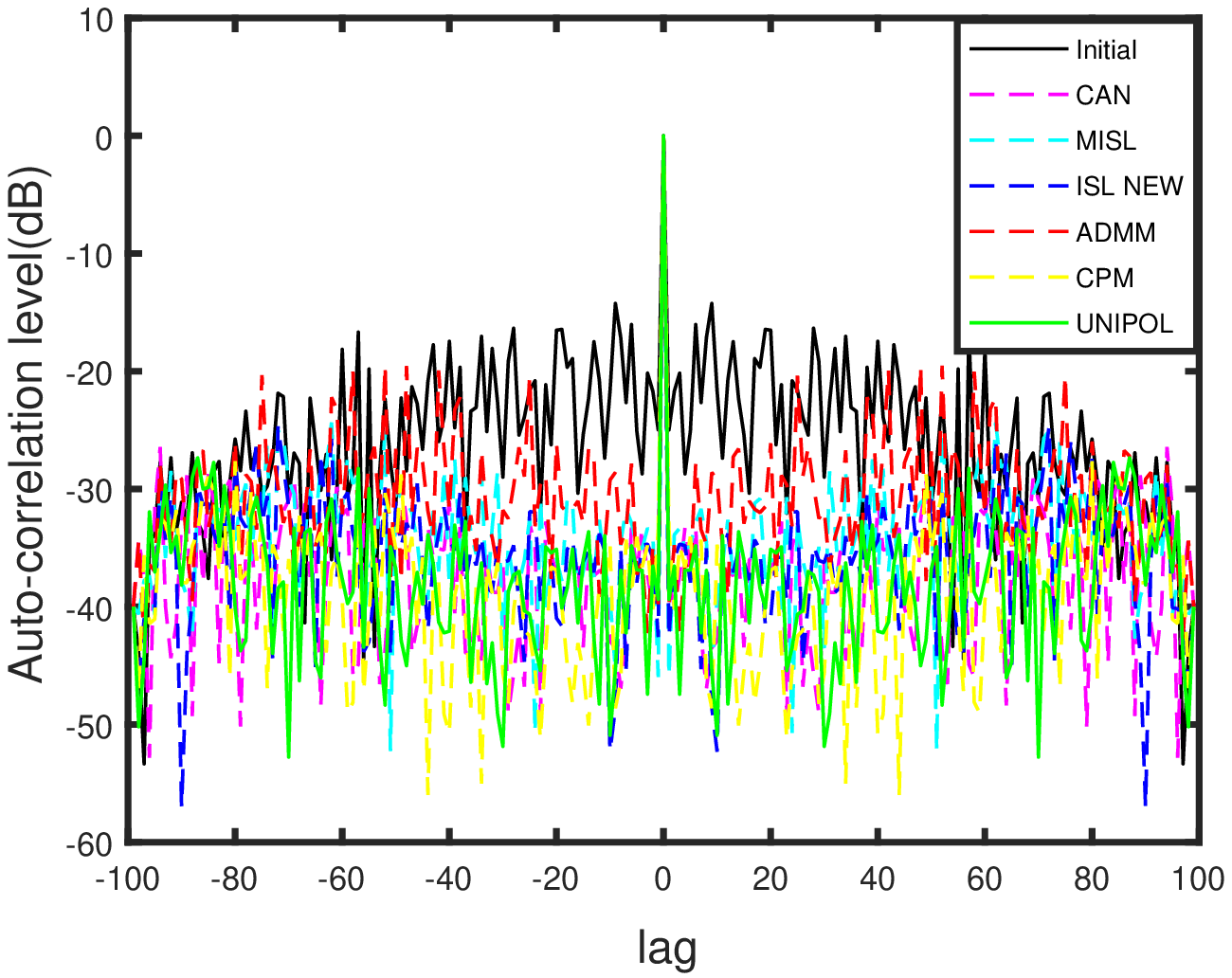}

}\subfloat[$20\log10\left(\left|r_{k}\right|/N\right)$ vs $k$ for $N=1000$]{\includegraphics[scale=0.55]{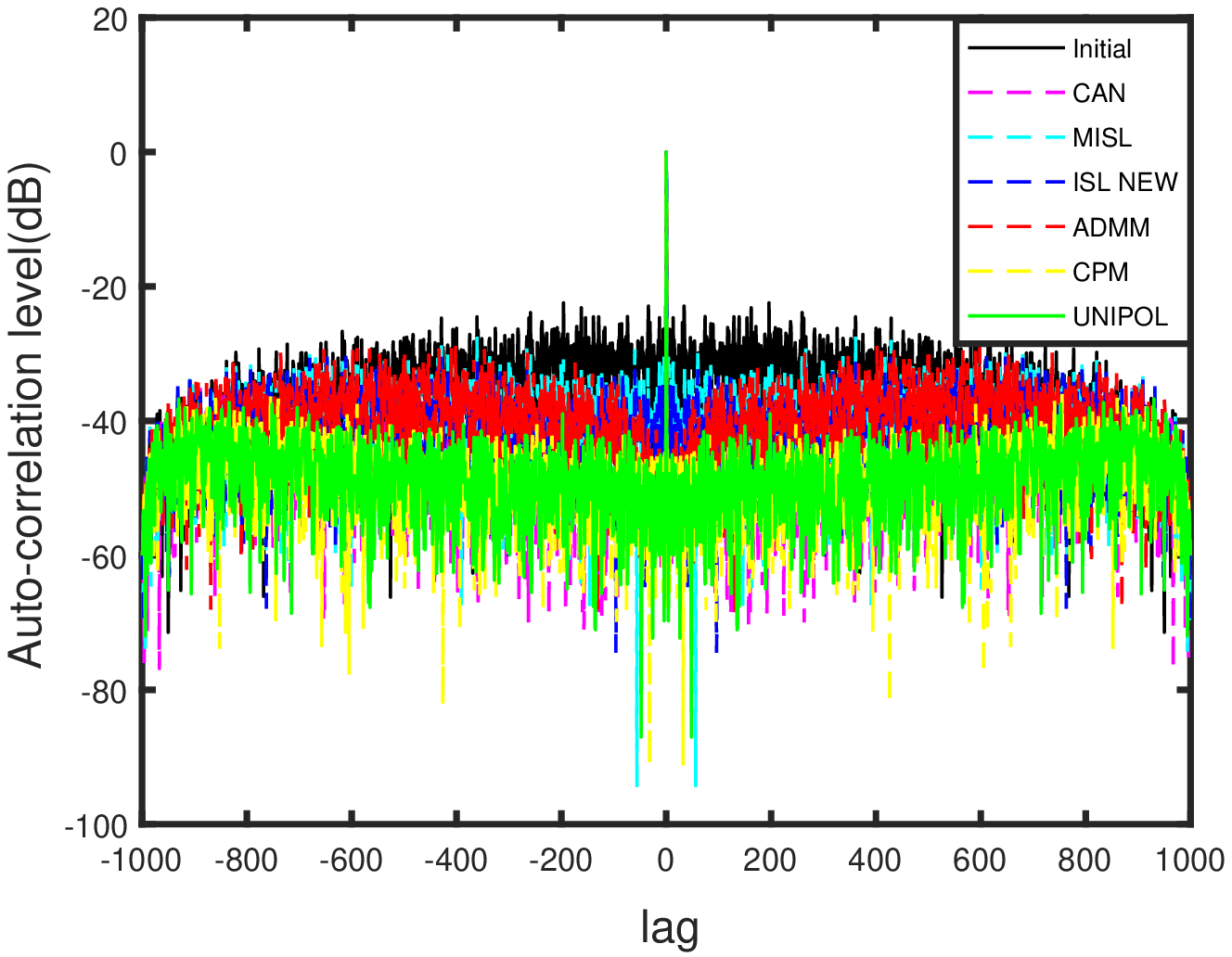}

}

\caption{UNIPOL simulations using the random initialization}
\end{figure}

CPM. We also implemented the SQUAREM accelerated MISL and ISL-NEW
algorithms. For better comparisons, all algorithms are initialized
using the same random sequence $i.e.,$ $e^{j\theta_{a}},a=1,..,N$,
where $\theta_{a}$ follows the uniform distribution between $\left[0,1\right]$.
The convergence criteria is selected as $1000$ iterations. 

Figure-1 consists of the ISL vs iterations, ISL vs time, and the autocorrelations
vs lag plots for two different sequence lengths $\left(N=100,1000\right)$.
From the simulation plots, we observe that all the algorithms has
started at the same initial point but converged to different minimum
values with different convergence rates. We noticed that in comparison
to the existing algorithms the UNIPOL has converged to a better minimum
in the least number of iterations (also in terms of CPU time). In
terms of the autocorrelation side-lobe levels, except the ADMM method,
all other methods have better sidelobe levels compared to initialization
sequences.

Figure-2 shows the comparison plots of different algorithms in terms
of average running time vs different sequence lengths. Here we didn't
consider the CAN and ADMM methods for comparisons because they both
are not actual ISL minimizers and to calculate the average running
time, for each length the experiment is repeated over $30$ Monte
Carlo runs. All the algorithms are run till they converge to a local
minima. From the simulation plots we observe that irrespective of
the sequence length the UNIPOL has taken the least time to design
sequences with better correlation properties.

\begin{figure}[ph]
\begin{centering}
\includegraphics[scale=0.62]{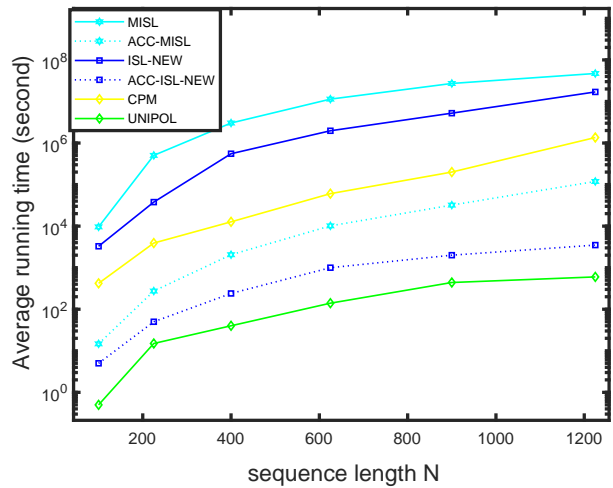}
\par\end{centering}
\caption{Average running time vs $N$}
\end{figure}

\section{{Conclusion}}

In this paper, we formulated the unimodular sequence design problem
as a constrained ISL minimization problem and proposed a majorization
minimization technique based algorithm named as UNIPOL where we solve
the series of polynomial optimization problems at every iteration.
The performance of the proposed algorithm is evaluated through the
numerical experiments and shows that it will perform well in terms
of the average convergence time.

\bibliographystyle{IEEEtran}
\bibliography{UNIPOL}

\end{document}